\documentstyle[12pt,preprint,aps]{revtex}

\def \to {\rightarrow}
\def \beq {\begin{equation}}
\def \eeq {\end{equation}}
\def \ba {\begin{eqnarray}}
\def \ea {\end{eqnarray}}
\def \jpsi {J/\psi}

\def \mtso {\langle 0 |{\cal O } ^{J/\psi}_{8} [ ^3S_1 ]| 0 \rangle}
\def \moso {\langle 0 |{\cal O } ^{\jpsi}_{8} [ ^1S_0 ]| 0 \rangle}

\def \mtpj {\langle 0 |{\cal O } ^{\jpsi}_{8} [ ^3P_J ]| 0 \rangle}

\begin{document}
\draft
\title{Diffractive $ \jpsi $ production through color-octet mechanism
in resolved photon processes at HERA}
\author{Jia-Sheng Xu}
\address{Department of Physics, Peking University, Beijing 100871, China}
\author{Hong-An Peng}
\address{ China Center of Advance Science and
Technology (World Laboratory), Beijing 100080, China \\
and Department of Physics, Peking University, Beijing 100871,
 China }
\maketitle

\begin{abstract}
We use the color-octet mechanism combined with the two gluon
exchange model for the diffractive $J/\psi$ production
in resolved photon processes. In the leading logarithmic approximation 
in QCD, we find that the diffractive $J/\psi$ production cross section
is related to the off-diagonal gluon density 
of the proton, the gluon density of the photon and to
the nonperturbative color-octet matrix element of $\mtso $. 
The cross section is found to be very sensitive 
to the gluon density of the photon.
As a result, this process may provide a wide window
for testing the two-gluon exchange model, studying the nature of hard
diffractive factorization breaking and may be particularly
useful in studying the gluon distribution of the photon.
And it may also be a golden place to test the color-octet mechanism
proposed by solving the $\psi'(J/\psi)$ surplus problem at the Tevatron.

\vskip 3mm
\end{abstract}

\vfill\eject\pagestyle{plain}\setcounter{page}{1}

\narrowtext

\par
In high-energy strong interactions the Regge trajectory with a vacuum quantum
number, the Pomeron, plays a particular and very important role in soft
processes in hadron-hadron collisions \cite{collins}.
However, the nature of the Pomeron and its interaction with hadrons
remain a mystery.
For a long time it had been understood that the dynamics of the
`` soft Pomeron '' was deeply tied to confinement.
However, it has been realized now that much can be learnt about hard
diffractive processes from QCD, which are now under study experimentally.
Of all these processes, the diffractive heavy quarkonium production
has drawn specially attention, because their large masses provide a natural
scale to guarantee the application of perturbative QCD.
In Refs.\cite{th1,th2}, the diffractive $\jpsi$ production
cross sections have been formulated in direct photoproduction and
deep inelastic scattering (DIS) processes in perturbative QCD.
In the framework of perturbative QCD the Pomeron is assumed to be
represented by a pair of gluons in the color-singlet state\cite{low}.
This two-gluon exchange model can successfully describe the experimental
results from HERA\cite{hera-ex}.
An important feature of this perturbative QCD model prediction is that
the cross section for the diffractive $\jpsi$ production is expressed
in terms of the square of the gluon density.
In a previous paper\cite{yuan}, we extend the idea of perturbative QCD
description of diffractive processes from $ep$ colliders to hadron colliders,
we give the formula for the diffractive $\jpsi$
production at hadron colliders in the leading logarithmic approximation (LLA)
in QCD by using the two-gluon exchange model.
We introduce the color-octet mechanism to realize the color-octet $c \bar c$
pair evolving into $\jpsi$ meson
and show the importance of diffractive $\jpsi$ production at
hadron colliders to the study of small $x$ physics,
the property of diffractive process, the nature of the Pomeron and even
for the test of color-octet production mechanism of heavy quarkonium.
In this paper, we discuss another diffractive process,
diffractive $\jpsi $ production through color-octet machnism
in resolved photon processes at HERA:
\beq
\gamma + P \to \jpsi + P + X    .
\eeq
This process is of special interesting  because the produced
$\jpsi $ is easy to be detected through its leptonic decay modes and
the diffractive $J/\psi$ production rate
is related to the off-diagonal gluon density 
in the proton and to the nonperturbative color-octet matrix element of
$J/\psi$, furthermore, the rate is found to be very sensitive 
to the gluon density of the photon. 
The measurement of this process at HERA  may provide a wide window
for studying the nature of diffraction, and may be particularly
useful in studying the gluon distribution of the photon.
And it may also be a golden place to test the color-octet mechanism
proposed by solving the $\psi'(J/\psi)$ surplus problem at the Tevatron.

\par
In quantum field theory, the photon is the gauge boson mediating the
electromagnetic interactions through the coupling to charged particles,
in this respect, the photon appeares to be an elementary point-like
particle. In other respect, the photon is subject to quantum fluctuation,
it can fluctuate into a quark-antiquark pair ($q{\bar q}$) plus all
other Fock states. If in
photon-hadron interactions, the fluctuation time is large copared to the
interaction time, the photon interacts with hadron through the interaction
of these Fock states with hadron. This is the hadron-like nature of the
photon which is well described by the vector dominance model (VDM)\cite{vdm}.
This dual nature of the photon have been studied in $ \gamma {\rm P}$ and 
$\gamma \gamma, \gamma ^* \gamma $ scattering processes \cite{review-ph}.
There are two basic types  of inclusive processes where the partonic
structure of the photon is investigated:
one is the deep inelastiv scattering $ e + \gamma \to e + X $ processes
( DIS$_{\gamma}$), where the structure function of the photon
${\rm F }_{2}^{\gamma} ( x, {\rm Q}^2 )$  is measured\cite{review-ph,f2-ph}.
the other is th large ${\rm P_T}$ jet and charged particle production
in $\gamma {\rm P}$ and $\gamma \gamma $ collisions, where individual
parton densities in the photon are
probed\cite{review-ph,amy-g,h1-g,zeus-g}.
The measurements of the photon structure function
${\rm F }_{2}^{\gamma} ( x, {\rm Q}^2 )$  in DIS$_{\gamma}$ are
directly sensitive to the quark structure of the photon, however
the presently available data on ${\rm F }_{2}^{\gamma} ( x, {\rm Q}^2 )$
are not precise enough to extracted the gluon distribution function of
the photon through QCD evolution studies.
First evidence for the gluon content of the photon was
shown by KEK TRISTAN experiments through the study of two-photon
production of large ${\rm P_T}$ jets\cite{amy-g}.
Recently,, the Leading order (LO) gluon destribution of the photon
in the fractional momentum rage $ 0.04 \leq x_{\gamma} \leq 1 $ at
the average factorization scales $75$ and $38 {\rm GeV}^2 $
have been extracted by H1 Collaboration using
dijet and large ${\rm P_T}$ charged particles photoproduction data\cite{h1-g}.
More processes should be studied, in order to obtain more
precise data to extracte the gluon content of the photon.
We will see through the following study that
diffractive $\jpsi $ production in resolved photon processes at HERA
can give valuable information about the gluon content of the photon.

\par
Now, we discuss diffractive $\jpsi $ production
in resolved photon processes at HERA in two-gluon exchange model (Fig. 1).
Within the nonrelativistic chromodynamics (NRQCD)
framework \cite{nrqcd,power} $\jpsi $ is described in terms
of Fock state decompositions as
\ba
|\jpsi \rangle &=& O(1)~ |c{\bar c}[^3S_{1}^{(1)}] \rangle +
            O(v) |c{\bar c}[^3P_{J}^{(8)}] g \rangle  \nonumber \\
        & & + O(v^2) |c{\bar c}[^1S_{0}^{(8)}] g \rangle +
            O(v^2) |c{\bar c}[^3S_{1}^{(1,8)}] g g \rangle  \nonumber \\
        & & + O(v^2) |c{\bar c}[^3P_{J}^{(1,8)}] g g\rangle + \cdots ,
\ea
where the $c{\bar c}$ pairs are indicated within the square brackets in
spectroscopic notation. The pairs' color states are indicate by singlet (1)
or octet (8) superscripts. The color octet $c{\bar c}$ states can make
a transition into a physical $\jpsi$ state by soft chromoelectric
dipole ($E1$) transition(s) or chromomagnitic dipole ($M1$) transition(s)
\beq
(c {\bar c})[^{2S + 1}L_{j}^{(8)}] \to \jpsi  .
\eeq
The color-octet contributions are essential for cancelling the logarithmic
infrared divergences which appear in the color-singlet model calculations of
the production cross sections and annihilation decay rates for P-wave
charmonia, and for solving the $\psi^{\prime}$
and direct $\jpsi$ ``surplus'' problems at the Fermilab
Tevatron\cite{cdfjpsi,brt-cho}.
The NRQCD factorization scheme \cite{nrqcdfs} has
been established to systematically separate these soft interactions from the
hard interactions.
In NRQCD, the long distance evolving process is described by the
nonperturbative matrix elements of four-fermion operators
and NRQCD power counting rules can be exploited to determine the
dominant matrix elements in various processes\cite{power}.
For $\jpsi $ production, the color-octet matrix elements,
$\mtso$ , $\moso$ and $ \mtpj $ are all scaling as $m_c^3 v_c^7 $. So these
color-octet contributions to $\jpsi $ production must be included for
consistency.
But in diffractive $\jpsi$ production, due to the nuture of vacuum quantum
number exchange, the color-octet $^1 S_0 $ and $^3 P_J$ subprocesses do
not contribute to the diffractive $\jpsi$ production.
So the diffractive $\jpsi$ production in resolved photon process in Fig. 1
can be realized via
the color-octet ${}^3S_1$ channel only, in which the $c\bar c$ pair in a
configuration of ${}^3S_1^{(8)}$ is produced in hard process as the
incident gluon interacts with the proton by $t$ channel color-singlet
exchange (the two-gluon ladder parametrized Pomeron),
and then evolve into the physical state $\jpsi$
through emitting soft gluons which carry little momentum and will cause
little change to the
final state $\jpsi$ spectrum in the diffractive processes.

\par
For the diffractive subprocesses, $gp\to c{\bar c}[^3S_{1}^{(1)}] p$,
the leading contribution
comes from the diagrams shown in Fig.2.
Due to the nuture of vacuum quantum
number exchange,
we know that the real part of the amplitude cancels out in the leading
logarithmic approximation.
The first two diagrams are similar to those calculated in direct
photoproduction process, and the rest diagrams are new due to the existence
of the gluon-gluon interaction vertex, those new diagrams are needed to
guarantee the gauge invariance.
For Fig.2(a), the imaginary part of the short distance amplitude
${\cal A}(g_a p\to (c\bar c)_b[^3S_1^{(8)}]p)$, to leading logarithmic
contribution, is\cite{yuan}
\beq
\label{ima}
{\rm Im}{\cal A}^{(a)}=F\times
       {{1\over 9}\delta^{ab}}
       \frac{1}{s} (s g_{\mu \nu} - 2 p_{2 \mu}
       q_{\nu} - 2 P^{\psi}_{\mu} p_{2 \nu} ) \varepsilon _{g}^{\mu }
       \varepsilon _{\psi}^{\nu } 
       \int\frac{dk_T^2}{k_T^4}\frac{1}{m_c^2}
       f(x^{\prime }, x^{\prime \prime}; k_T^2) ,
\eeq
where $F=\frac{3\pi}{2} g_s^3 m_c s$,
$a$ and $b$ are the color indexes of the incident gluon and
the $c\bar c$ pair in color-octet ${}^3S_1$ state.
Factor $1\over 9$ is the color factor.
$p_2, q $ and $ P^{\psi} $ are the four-momenta of the proton, the gluon
from the photon and the $\jpsi$ respectivly, $K_T$ is the transverse
momentum of the gluon attached with the proton. 
$s = ( q + p_2)^2 $.   $\varepsilon _{g}^{\mu }$ and
$\varepsilon _{\psi}^{\mu }$ are the polarization vectors
of the gluon from the photon and $\jpsi $.
The function $ f(x^{\prime }, x^{\prime \prime}; k_T^2)$ is related to
the so-called off-diagonal gluon distribution function
$G(x^{\prime }, x^{\prime \prime}; k_T^2)$ \cite{ji} by

\beq
f(x^{\prime}, x^{\prime\prime};k_T^2) =
\frac{\partial G(x^{\prime}, x^{\prime\prime}; k_T^2)}
{\partial {\rm ln} k_T^2}
\eeq
Here, $x^{\prime }$ and $x^{\prime\prime}$ are the momentum fraction of the
proton  carried by the two gluons.
In our calculations, we set the momentum transfer  of the proton
$t$ equal to zero, i.e., $t=(k-k^{\prime})^2=0$.

\par
For Fig.2(b), the result is,
\beq
\label{imb}
{\rm Im}{\cal A}^{(b)}=-F\times({-{1\over 72} \delta^{ab}})
       \frac{1}{s} (s g_{\mu \nu} - 2 p_{2 \mu}
       q_{\nu} - 2 P^{\psi}_{\mu} p_{2 \nu} ) \varepsilon _{g}^{\mu }
       \varepsilon _{\psi}^{\nu } 
       \int\frac{dk_T^2}{k_T^4}\frac{1}{m_c^2+k_T^2}
       f(x^{\prime }, x^{\prime \prime}; k_T^2) ,
\eeq
where the color factor is $- {1\over 72}$.
Unlike the case of the diffractive photoproduction processes, the color
factors of these two diagrams (Fig.2(a) and Fig.2(b)) are not the same.
The leading part of the contributions from these two diagrams (which is
proportional to $1\over k_T^4$) can not cancel out each other.
After integrating the loop momentum $k$, for small $k_T$ this will lead
to a linear
singularity, not a logarithmic singularity (proper in QCD) as that in
diffractive direct photoproduction process \cite{th1}.
So, there must be some other diagrams to cancel out the leading part of
Fig.2(a) and Fig.2(b) to obtain the correct result.
This is also due to the gauge invariance requirement.
As mentioned above, in QCD due to the nonabelian $SU(3)$ gauge theory
there are additional diagrams shown in Fig.2(c)-(e) as compared
with that in direct photoproduction at $ep$ colliders.
By summing up all these diagrams together, we expect that in the final
result the leading part singularity which is
proportional to $1\over k_T^4$ will be canceled out,
and only the terms proportional to $1\over k_T^2$ will be retained.

The contribution from Fig.2(c) is,
\beq
\label{imc}
{\rm Im}{\cal A}^{(c)}= 4 F\times ({-{1\over 8}\delta^{ab}})
       \frac{1}{s} (s g_{\mu \nu} - 2 p_{2 \mu}
       q_{\nu} - 2 P^{\psi}_{\mu} p_{2 \nu} ) \varepsilon _{g}^{\mu }
       \varepsilon _{\psi}^{\nu } 
       \int\frac{dk_T^2}{k_T^4}\frac{1}{4m_c^2+k_T^2}
       f(x^{\prime }, x^{\prime \prime}; k_T^2) ,
\eeq
where $-{1\over 8}$ is color factor.
When we perform the integral over the loop momentum $k$, the main large
logarithmic contribution comes from the region
${1\over R_N^2}\ll k_T^2\ll M_\psi^2$ ($R_N$ is the nucleon
radius)\cite{th1}. So, we calculate the amplitude as an expansion of $k_T^2$.
From Eqs.(\ref{ima}), (\ref{imb}), (\ref{imc}),
we can see that the leading part singularity from Fig.2(a)-(c) are
canceled out as expected.

By the same reason, for Fig.2(d) and Fig.2(e), the leading part of each
diagram is proportional to $1\over k_T^4$. However, their sum is only
proportional to $1\over k_T^2$ because the leading part is canceled out.
Their final results is,
\beq
{\rm Im}{\cal A}^{(d + e)}=F\times ({-{1\over 2}\delta^{ab}})
       \frac{1}{s} (s g_{\mu \nu} - 2 p_{2 \mu}
       q_{\nu} - 2 P^{\psi}_{\mu} p_{2 \nu} ) \varepsilon _{g}^{\mu }
       \varepsilon _{\psi}^{\nu } 
       \int\frac{dk_T^2}{k_T^2}\frac{2}{16m_c^4}
       f(x^{\prime }, x^{\prime \prime}; k_T^2) .
\eeq

Adding all the contributions from Fig.2(a)-(e), we get
the imaginary part of the short distance amplitude,
\ba
{\rm Im}{\cal A}(gp\to (c\bar c)[^3S_1^{(8)}]p)
 &=& F\times {(-{13\over 18}\delta^{ab})}
  \frac{1}{s} (s g_{\mu \nu} - 2 p_{2 \mu}
  q_{\nu} - 2 P^{\psi}_{\mu} p_{2 \nu} ) \varepsilon _{g}^{\mu }
  \varepsilon _{\psi}^{\nu }  \nonumber    \\
 &~& \times \int\frac{dk_T^2}{k_T^2}\frac{1}{M_\psi^4}
  f(x^{\prime},x^{\prime\prime};k_T^2).
\ea
Where $M_{\psi} = 2 m_c $ are used.
It is expected that for small $x$, there is no big difference between
the off-diagonal and
the usual diagonal gluon densities\cite{off-diag}.
So, in the following calculations, we estimate the production rate by
approximating the off-diagonal gluon density by 
the usual diagonal gluon density, 
$G(x^{\prime},x^{\prime\prime};Q^2)\approx xg(x,Q^2)$,
where $x=M_{\psi}^{2}/s$.
Finally, in the leading logarithmic approximation (LLA), we obtain
the imaginary part of the short distance amplitude,
\beq
\label{e3}
{\rm Im}{\cal A}(gp\to (c\bar c)[{}^3S_1^{(8)}]p)=
{(-{13\over 18}\delta^{ab})}\frac{F}{M_\psi^4}xg(x,\bar{Q}^2),
  \frac{1}{s} (s g_{\mu \nu} - 2 p_{2 \mu}
  q_{\nu} - 2 P^{\psi}_{\mu} p_{2 \nu} ) \varepsilon _{g}^{\mu }
  \varepsilon _{\psi}^{\nu } 
\eeq

\par
Using Eq.(\ref{e3}), we get the cross section
for the diffractive subprocess $gp\to \jpsi p$,
\beq
\label{e4}
\frac{d\hat\sigma(gp\to \jpsi p)}{dt}|_{t=0}
    =   \frac{169\pi^4m_c}{32\times 27}\frac{\alpha_s(\bar{Q}^2)^3\mtso }
        {M_\psi^8} [xg(x,\bar{Q}^2)]^2.        
\eeq
We have also calculated the color-octet $^1S_0$ and $^3P_J$ subprocesses.
As expected they do not contribute to the diffractive $\jpsi$ production.
That is, only the color-octet $^3S_1$ contributes in this
process. So the diffractive $\jpsi$ production considered in this paper
is sensitive to the matrix element $\mtso $, which
is very important to describe the prompt $\jpsi$ production at the
Tevatron.
Therefore, the diffractive $\jpsi$ production at hadron colliders
would provide a golden place to test the color-octet mechanism in the
heavy quarkonium production.

Provided the partonic cross section Eq.(\ref{e4}) above, we can get the cross
section of diffractive $\jpsi$ production in resolved photon processes.
Assuming hard diffractive factorization, the differential cross section can
be expressed as
\ba
\label{e5}
\frac{d\sigma(\gamma + P \to \jpsi + P + X )}{dt}|_{t=0} & = &
       \frac{169\pi^4m_c}{32\times 27}\frac{\alpha_s(\bar{Q}^2)^3\mtso}
        {M_\psi^8}  \nonumber  \\
    &~& \times \int_{x_{1\rm min}}^1 dx_1g(x_1,\bar Q^2)[xg(x,\bar{Q}^2)]^2,
\ea
and
\beq
\label{total}
\sigma(\gamma + P \to \jpsi + P + X ) =
\frac{1}{b} \frac{d\sigma(\gamma + P \to \jpsi + P + X )}{dt}|_{t=0}
\eeq
where $ b $ is the $ t $ slope of the $\jpsi $ diffractive production in
resolved photon processes, $x_1$ is the longitude momentum fraction of the
photon  carried by the incident gluon.
So, the c.m. energy of the gluon-proton
system is $s=x_1 S_{\gamma P} $, where $S_{\gamma P}$ is the total
c.m. energy of the photon and proton system.
Then, $x=M_\psi^2/s=M_\psi^2/(x_1 S_{\gamma P})$.

\par
For numberical predictions,
we use $m_c = 1.5~{\rm GeV}, \Lambda_4 = 235~ {\rm MeV} $, and set the
factoriztion scale and the renormalization scale both equal to the
mass of $\jpsi$, {\it i.e.},
${\bar Q}^2 = M_{\psi}^{2} $. We assume the $ t $ slope $b$ is the same as
the value in the direct photoproduction of $\jpsi $ process, i.e.,
$ b = 4.5 {\rm GeV}^{- 2} $\cite{hera-ex}.
For the color-octet matrix elements
$\mtso $ we use the values determined by
Beneke and Kr$\ddot{a}$mer \cite{beneke} from fitting the direct
$\jpsi$ production data at the Tevatron \cite{cdfjpsi} using GRV LO parton
distribution functions\cite{grv}:
\beq
$$ \mtso = 1.12 \times 10^{-2} ~{\rm GeV}^3 ,$$  
\eeq
As usual, in order to suppress the Reggon contributions, we set
$x \leq 0.05$, so $x_{ {1\rm min} } = 20 M_{\psi}^2/S_{\gamma P} $.

In Fig. 3, we plot the total cross section
$\sigma(\gamma + {\rm p} \to \jpsi + {\rm p} + X )$ as a function of the total
c.m. energy of the photon and proton system
${\rm Ecm} (\gamma {\rm p})$
in the range $160 \leq {\rm Ecm} (\gamma {\rm p}) \leq 250 ~{\rm GeV} $.
We use the GRV LO gluon distribution function for the proton\cite{grv},
and GRV LO, SaS2M, and WHIT1 parametrizations for the gluon distribution
of the photon\cite{grv-ph,sas,whit}.  The solid curve is for GRV LO,
dotted for WHIT1, dashed for SaS2M  parametrizations for the gluon
distribution of the photon.
The results calculating from the GRV LO and WHIT1 gluon
distribution sets are almost the same, while the cross section is
larger for SaS2M set.
In the ${\rm Ecm} (\gamma {\rm p})$ range
considered, the total cross section is in the region
$0.24 ~{\rm nb} < \sigma(\gamma + {\rm p} \to \jpsi + {\rm p} + X ) < 0.73
~{\rm nb} $. So this process can be studied at DESY HERA with present
integrated luminosity.
In Fig. 4, we plot the cross section
$\sigma(\gamma + {\rm p} \to \jpsi + {\rm p} + X )$ as a function of
the lower bound of $ x_1$ in the integral of Eq.(\ref{e5})
at $ {\rm Ecm} (\gamma {\rm p}) = 200 ~{\rm GeV}$ using  the GRV LO gluon
distribution function for the photon and proton. From this figure, we
see that the main contribution to the total cross section comes from
the region $ 10^{-2} < x_1 < 5.0 \times 10^{-1} $, which contributes
$ 89\% $ of the total cross section. The region of
$ 10^{-2} < x_1 < 5.0 \times 10^{-1} $ corresponds to the region
$ 4.8 \times 10^{-3} < x < 2.4 \times 10^{-2} $  where the gluon desity of the
proton is well determined by HERA experiments. So the
measurement of diffractive $\jpsi $ production in resolved photon
process at HERA can determine the gluon desity of the photon
in the region of $ 10^{-2} < x_1 < 5.0 \times 10^{-1} $.

\par
In the above, we have assumed hard diffractive factorization in the
resolved photon process.
Recently, a factorization theorem
has been proven by Collins \cite{jcollins} for the lepton induced
hard diffractive scattering processes,
such as diffractive deep inelastic scattering (DDIS) and diffractive
direct photoproduction of jets.
In contrast, no factorization theorem has been established for hard
diffraction in resolved photon processes and hadron-hadron collisions.
At large $|t|$ ($t$ is the square of the hadron's four-momentum transfer),
where perturbative QCD applies to the Pomeron, it has been proven that
there is a leading twist contribution which breaks the factorization
theorem for hard diffraction in hadron-hadron collision \cite{copom}.
This coherent hard diffraction was observed by the UA8 Collaboration in
diffractive jet production, in this experiment  $t$ is in the region
$ -2 \leq t \leq -1 $ GeV$^2$ \cite{ua8}.
In phenomenology,
the large discrepancy between the theoretical prediction and
the Tevatron date on the diffractive production of
jets and weak bosons, {\it at al.}, signals
a breakdown of hard diffractive factorization in hadron-hadron
collisions \cite{testa}.
Since in resolved photon processes,
the photon behaves likes a hadron, we expect there are
nonfactorization
effects in diffractive resolved photon processes,
but the nature of hard diffrative factorization breaking is unclear.
Goulianos have concluded that the
breakdown of hard diffractive factorization in hadron-hadron collisions is
due to the breakdown of the Regge factorization already observed in soft
diffraction\cite{gouln2} and this effects can indicated by  the so-called
renormalization factor D  which takes the role of the
survival probability for hadron emerges from the diffractive collision
intact\cite{goulianos,gouln2,soper}:

\beq
\label{dfactor}
D = {\rm min} (1, \frac{1}{N}) ,
\eeq
with
\beq
\label{nv}
N = \int_{\xi_{{\rm min} }}^{\xi_{{\rm max}}} d \xi \int^{0}_{- \infty} d t
f_{{\rm I\!P}/p }(\xi,t) ,
\eeq
where $\xi_{{\rm min}} = M_0^2/Ecm $ with $M_0^2 = 1.5 {\rm GeV}^2 $
(effective threshold) and $\xi_{{\rm max}} = 0.1 $(coherence limit) and
$f_{{\rm I\!P}/p }(\xi,t)$ is the Pomeron flux factor.
For our case, ${\rm Ecm} (\gamma {\rm p})$
in the range $160 \leq {\rm Ecm} (\gamma {\rm p}) \leq 250 ~{\rm GeV} $,
the renormalization factor ${\rm D(Ecm)}$in the range
$ 1/3 > D > 1/4$. The total cross section shown in Fig. 3 should
multiplied by the renormalization factor ${\rm D(Ecm)}$ provided
one takes into nonfactorization effects in this way. But the other results
are unchanged. So the
measurement of diffractive $\jpsi $ production in resolved photon
process at HERA can shed light on the nature of the hard diffractive
factorization breaking.

\par
In conclusion, in this paper 
We use the color-octet mechanism combined with the two gluon
exchange model for the diffractive $J/\psi$ production
in resolved photon processes. In the leading logarithmic approximation 
in QCD, we find that the diffractive $J/\psi$ production cross section
is related to the off-diagonal gluon density 
in the proton, the gluon density of the photon and to
the nonperturbative color-octet matrix element of $J/\psi$. 
The cross section is found to be very sensitive 
to the gluon density of the photon.
Since the color-octet $^1S_0$ and $^3P_J$ subprocesses
do not contribute to the diffractive $\jpsi$ production,
that is, only the color-octet $^3S_1$ contributes in this
process. So the diffractive $\jpsi$ production considered in this paper
is sensitive to the matrix element $\mtso $, which
is well determined from  the direct $\jpsi$ production data at the
Tevatron.
Therefore, the diffractive $\jpsi$ production in resolved photon processes
would provide a golden place to test the color-octet mechanism in the
heavy quarkonium production. Furthermore
the measurement of diffractive $\jpsi $ production in resolved photon
process at HERA can shed light on the nature of the hard diffractive
factorization breaking.

\begin{center}
{\bf\large Acknowledgments}
\end{center}
This work is supported  by the National Natural Science Foundation of
China, Doctoral Program Foundation of Institution of Higher Education of
China and Hebei Natural Province Science Foundation, China.

\newpage
\centerline{\bf \large Figure Captions}
\vskip 2cm
\noindent
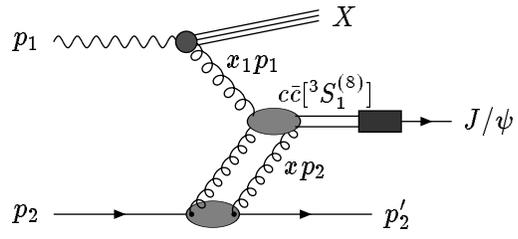
Fig.1. Sketch diagram for the diffractive $J/\psi$ production
in resolved photon processes 
in perturbative QCD. The black box represents the long distance process
for color-octet $c\bar c$ pair in $^3S_1$ state evolving into $\jpsi$.

\noindent
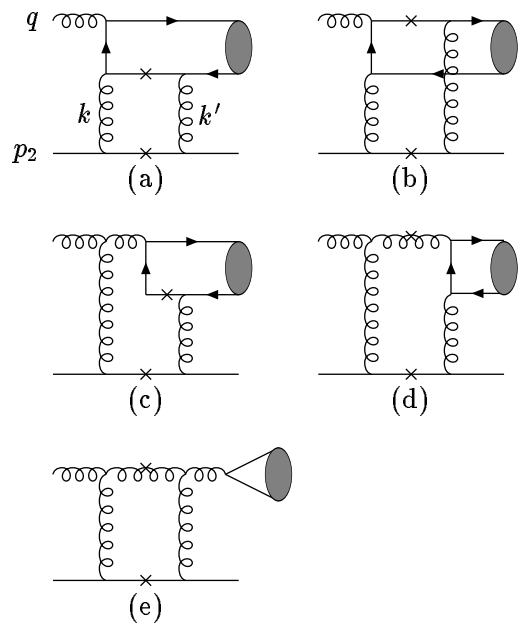
Fig.2. The lowest order perturbative QCD diagrams for the diffractive
$\jpsi$ production in resolved photon processes. The ``$\times $ ''
represents cutting the line.

\noindent
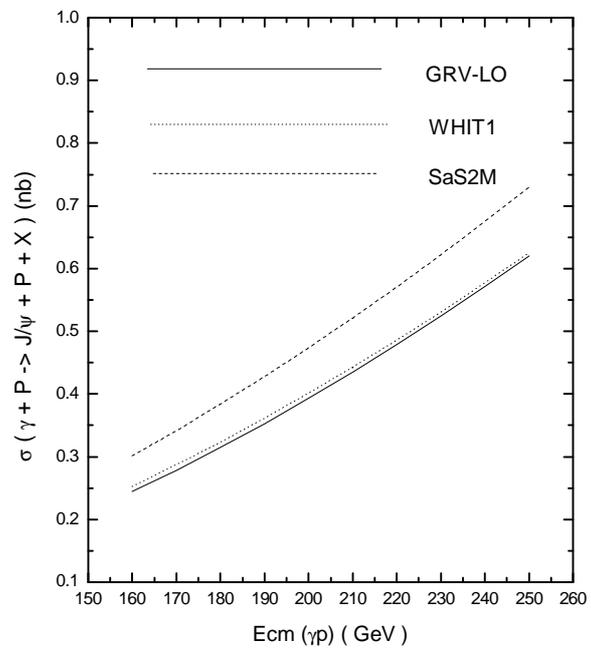
Fig.3. The total cross section
$\sigma(\gamma + {\rm p} \to \jpsi + {\rm p} + X )$ as a function of the
total c.m. energy of the photon and proton system
${\rm Ecm} (\gamma {\rm p})$
in the range $160 \leq {\rm Ecm} (\gamma {\rm p}) \leq 250 ~{\rm GeV} $.
The solid curve is for GRV LO,
dotted for WHIT1, dashed for SaS2M  parametrizations for the gluon
distribution of the photon.

\noindent
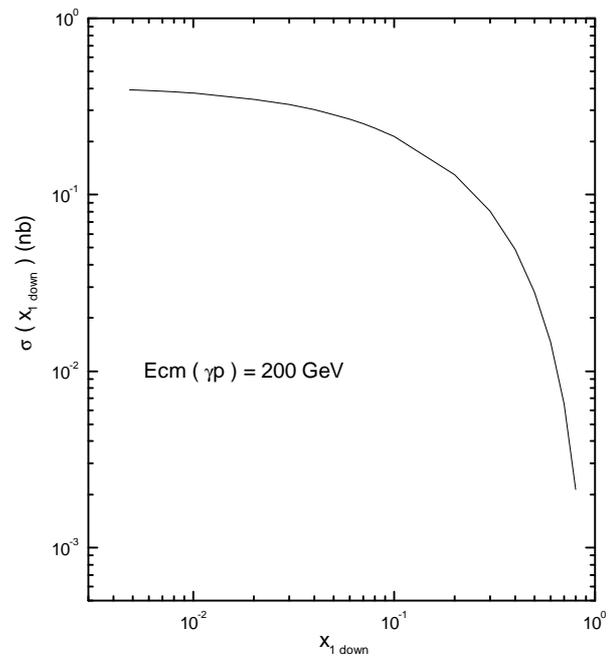
Fig.4. The total cross section
$\sigma(\gamma + {\rm p} \to \jpsi + {\rm p} + X )$ at the
total c.m. energy of the photon and proton system
${\rm Ecm} (\gamma {\rm p}) = 200 ~{\rm GeV}$
as a function of the lower bound of $x_1$ in the integral of
Eq.(\ref{e5}) using GRV LO
parametrization for the gluon
distribution of the photon.


\end{document}